\documentclass[sigconf]{acmart}

\usepackage{xspace}
\usepackage{array}

\newcommand{\Kellen}{\textsc{Kellen}\xspace}
\newcommand{\Wendy}{\textsc{Wendy}\xspace}
\newcommand{\David}{\textsc{Anonymous}\xspace}

\newcommand{\Lisa}{\textsc{Lisa}\xspace}
\newcommand{\Andy}{\textsc{Anonymous}\xspace}

\newcommand{\Abigail}{\textsc{Anonymous}\xspace}

\newcommand{\Tim}{\textsc{Tim}\xspace}

\newcommand{\Tom}{\textsc{Tom}\xspace}

\newcommand{\BUGSS}{\textsc{BUGSS}\xspace}
\newcommand{\Cambridge}{\textsc{Cambridge Biomakespace}\xspace}
\newcommand{\CCL}{\textsc{Counter Culture Labs}\xspace}

\newcommand{\SoundBio}{\textsc{SoundBio Lab}\xspace}
\newcommand{\HiveBiolab}{\textsc{Hive Biolab}\xspace}
\newcommand{\OSN}{\textsc{Open Science Network}\xspace}

\newcommand{\Spira}{\textsc{Spira}\xspace}

\AtBeginDocument{%
  \providecommand\BibTeX{{%
    \normalfont B\kern-0.5em{\scshape i\kern-0.25em b}\kern-0.8em\TeX}}}

\setcopyright{iw3c2w3}
\copyrightyear{2022}
\acmYear{2022}
\acmDOI{10.1145/1122445.1122456}

\acmConference[DIS '22]{Designing Interactive Systems 2022}{June 03--05, 2022}{Virtual}
\acmBooktitle{Designing Interactive Systems 2022}
\acmPrice{15.00}
\acmISBN{978-1-4503-XXXX-X/18/06}

\begin{document}

\title[``Short on time and big on ideas'': Perspectives from Lab Members on DIYBio Work]{``Short on time and big on ideas'': Perspectives from Lab Members on DIYBio Work in Community Biolabs}





\author{Orlando de Lange}
\affiliation{%
  \institution{University of Washington}
  \city{Seattle, WA}
  \country{USA}}
\email{odl@uw.edu}

\author{Kellie Dunn}
\affiliation{%
  \institution{University of Washington}
  \city{Seattle, WA}
  \country{USA}}
  \email{kelliead@uw.edu}

\author{Nadya Peek}
\affiliation{%
  \institution{University of Washington}
  \city{Seattle, WA}
  \country{USA}}
\email{nadya@uw.edu}


\begin{abstract}

DIYbio challenges the status quo by positioning laboratory biology work outside of traditional institutions. 
HCI has increasingly explored the DIYbio movement, but we lack insight into sites of practice such as community biolabs.
Therefore, we gathered data on eleven community biolabs by interviewing sixteen lab managers and members.
These labs represent half of identified organizations in scope worldwide. 
Participants detailed their practices and motivations, outlining the constraints and opportunities of their community biolabs. 
We found that lab members conducted technically challenging project work with access to high-end equipment and professional expertise. 
We found that the unique nature of biowork exacerbated challenges for cooperative work, partially due to the particular time sensitivities of work with living organisms.
Building on our findings, we discuss how community biolab members are creating new approaches to laboratory biology and how this has design implications for systems that support non-traditional settings for scientific practice.

\end{abstract}



\begin{teaserfigure}
  \includegraphics[width=\textwidth]{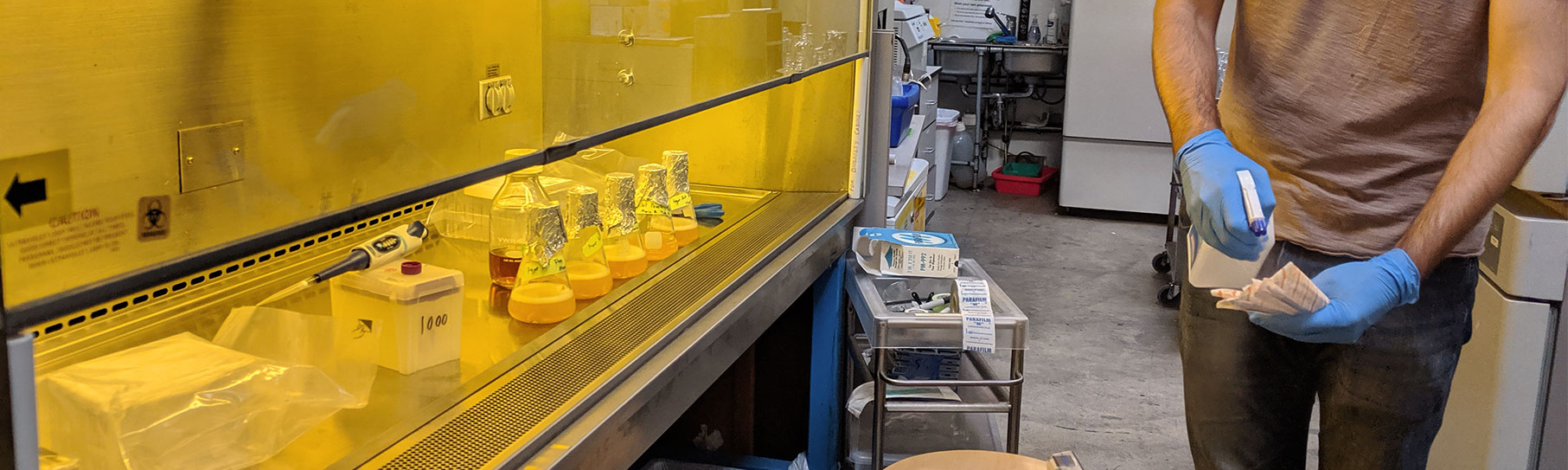}
  \caption{Biosafety cabinet, pipette, and materials for a DIYbio project in a community biolab.}
  \Description[A person stands next to a biosafety cabinet in a community biolab.]{The biosafety cabinet is wide and has a glass front. Inside of the cabinet is a pipette tool and containers of biological materials. Next to the cabinet, a person wearing blue disposable gloves is holding a spray bottle.}
  \label{fig:teaser}
\end{teaserfigure}

\maketitle

\section{Introduction}

Professional laboratory biology research is is a powerful system for generating novel technologies, though one that some have argued will inevitably be used to further the destructive, short-term interests of those in positions of power \cite{10.1105/tpc.2.4.275}. Donna Haraway, in discussion with synthetic biologist Drew Endy \cite{Haraway2019Tools}, entertains the idea that those wishing to create an alternative, more socially just, technoscientific future may find hope in throwing their resources behind those who are \textit{"staying with the trouble"}. 
Rather than attempting to shut biotechnology down \cite{10.1105/tpc.2.4.275}, she argues for strengthening the communities who are \textit{"telling better stories"}.
She offers, \textit{"Biology is at least two different things: it’s the historical discourse of biological knowledge making, and it’s the biological world, which isn’t the same thing. [...]
this [biological] knowledge-making apparatus, which tells a story about itself, is separating nature from culture, studying nature separately from human politics and religion and the rest. I know there’s a grain of truth in that story, but there’s also a huge amount of self-deception because the history of biology is the history of human beings engaging the world with technologies and the projects of knowing."}
Instead of separating biology from its cultural context, she argues for understanding them as inextricably linked \cite{doi:10.1086/703377}.

Biological knowledge-making is dominated by professionals working within siloed institutions. 
The barriers to laboratory biology work are particularly high, since practitioners require a dedicated space and specialist equipment and materials, including organisms and biomaterials, in addition to  technical expertise. 
However, \emph{community biolabs}, the subject of this work, have emerged as sites where individuals are invited to engage in work with the world of organisms and biomaterials, lowering the barriers by providing space and equipment. We are interested in how community biolabs are taking a new look at the social possibilities of laboratory biology knowledge, methods, and equipment. 

Community biolabs are independent not-for-profit spaces for biology work \cite{Kean2011-bs}, as well as being social organizations that support the users of the spaces as well as, in most cases, broader education and outreach work \cite{Scheifele2016-tk}. Each community biolab is independent and unique, though they are generally considered in association with the wider DIYbio movement \cite{Landrain2013}, and they have often been referred to as "DIYbio labs" \cite{Landrain2013}. We have chosen the term "community biolab" for this paper in the absence of any one dominant name, though we note that "community lab", "biohackspace" and "biomakerspace" are all additional terms used within and outside these various organizations. 

DIYbio is a term popularized by a group of open biotechnology advocates and community organizers in the Boston, MA area in 2008 \cite{Tocchetti2014-bh} and promoted via the DIYbio.org website and forum \cite{Landrain2013,Meyer2020DIYbioHistory}. 
Since then the term has been used far more broadly, often interchangeably or in conversation with the term biohacker \cite{MeyerBiohacking}.  In the years after the launch of DIYbio.org in 2008, community biolab spaces were launched in the US and internationally \cite{Landrain2013,DIYbiosphere}. 
DIYbio has been suggested, among other things, as a movement to forge ``bottom-up'' \cite{Keulartz2016} pathways to knowledge production and technology creation, as well as a movement to critique institutional technoscience \cite{Kera2017}, or both at once \cite{doi:10.1177/0162243915589634}. 
Unlike institutional technoscience, DIYbio tends to embrace the aesthetic qualities of laboratory biology practices, recognizing the meaning and value these qualities bring to biological knowledge making and blurring the lines between bioscience and bioart  \cite{Keulartz2016,Wilbanks2017}.

\citet{doi:10.1177/0162243915595091} argue that the "hacker mind-set" has extended from computer software to new domains, including electronics, hardware, and biology.
They argue that "hackerspaces" and "makerspaces" are places of emancipatory promise: where tools and technology can be reverse engineered and repurposed in the name of open source hardware or DIYbio.
We value critiques of maker culture \cite{ 10.1145/3274287,10.1145/2559206.2579405} and critiques of the repurposing of maker culture by institutions and governments
\cite{doi:10.1177/0162243915590861}. These lines of critique should and have been extended to DIYbio — \citet{Keulartz2016} explore the ethical ambivalence within DIYbio, including its potential to challenge but also to reinforce the norms of "BIG bio" and the "commodification of all aspects of life".
However, despite valid criticism, we also believe that makerspaces and community biolabs can provide "utopian glimmers" of collaborations and partnerships that do not uphold a problematic status quo
\cite{10.1145/2858036.2858506}.
In addition, we argue that the material practice of biological knowledge-making is markedly different than that of digital fabrication and hardware manufacturing.
Therefore, we see value in studying community biolabs, and furthermore understanding distinctions between community biolabs and makerspaces.

\citet{martin2015promise} argues that the maker movement comprises 1.) tools such as rapid prototyping equipment and microcontroller platforms 2.) community infrastructure such as in-person spaces and events and 3.) a mindset of playfulness and exploration.
This definition of the maker movement prioritizes making \emph{things} over making knowledge.
We believe we need a more in-depth understanding of how people are taking on projects of biological knowledge-making, and therefore seek to understand the specifics of DIYbio and the spaces in which it is practiced.
Furthermore, this definition of the maker movement includes the myriad makerspaces that are part of academic institutions or corporate labs.
Within DIYbio practice, the dominant sites are \emph{non}-institutional settings such as independent community biolabs (that are the focus of this study) or home labs and artists' studios.

\begin{figure*}[htbp]
\centering
\includegraphics[width=1.0\textwidth]{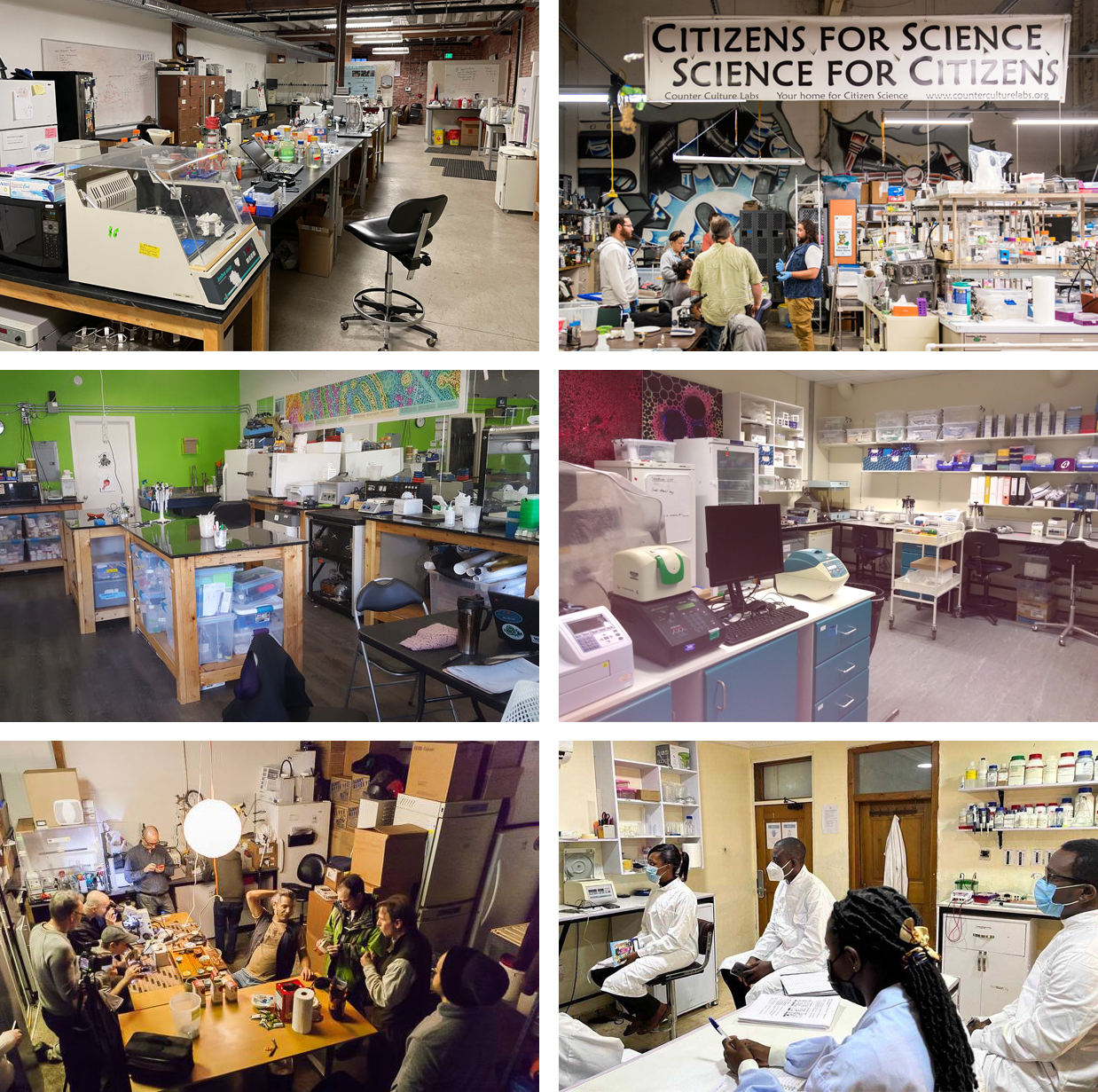}
\caption{Six of the eleven community bio labs included in our survey. Each of these lab spaces show setups typical of molecular biology laboratory practice, including workbenches, specialized equipment, and organized storage for chemicals and materials. While they range in size from a warehouse to a small corner of a room, they are overall smaller than comparable makerspaces, reflecting that the scale of work performed is often at the micro- and nano scale.
Clockwise from top left: \BUGSS, \CCL, \Cambridge, \HiveBiolab, \OSN, \SoundBio.}
\Description[A collage of six photos of community biolab spaces.]{On the top left: a basement room with equipment on long work benches. Top right: four people are meeting in a large space with lots of equipment under a banner that reads "Citizens for Science, Science for Citizens". Middle right: a corner of a lab with shelves on the walls and workbenches. Bottom right: four people wearing lab coats are having a meeting in a small room with tidy equipment and supplies. Bottom left: a group of people work gather around a table with project work, surrounded by shelving and boxes. Middle left: a brightly painted room with several work benches storing equipment and supplies.}
\label{fig:labs}
\end{figure*}

DIYbio has been highlighted as not only a productive subject of study, but also as a new area of HCI research practice  \cite{Fernando2016-tt}. 
The specific DIYbio practices of bio art \cite{Asgarali-Hoffman2021-rc, Hamidi2021-wv, Kuznetsov2012-kt} and open source laboratory hardware \cite{Fernando2020-bq} have been explored in HCI research. 
Microbes themselves have been promoted as novel design materials \cite{Barati2021-vb} or as living computers with which HCI researchers should be engaged in developing new conceptions of---and tools for---active design \cite{Pataranutaporn2020-qw}. 
We are interested in supporting the exploration of the possible intersections of HCI and DIYbio. We argue that a grounded understanding of the places in which DIYbio practitioners are already working and the challenges and opportunities of work in those spaces can inform future HCI research. 

The practices encompassed within DIYbio draw largely from those of biology research laboratories. Such laboratories tend to be technology-dense sites with workflows requiring materials to be moved between multiple computer-controlled machines as well as relying on computational data analysis and software tools to facilitate project planning and management. Novel approaches to the design of research tools for biology laboratories have been explored in the context of HCI \cite{Tabard2012,Mackay2002,Hu2015,Hincapie-Ramos2010,Hincapie-Ramos2011}. 
In their 2012 paper Kuznetsov and colleagues \cite{Kuznetsov2012-kt} lay out a vision for systems design research that engages with DIYbio from a range of stances. They suggest that novel interactive systems within DIYbio contexts can both provide new opportunities for project work while also being a material exploration of what DIYbio means as a technosocial movement. We argue that a grounded understanding of the places in which DIYbio practitioners are already working and the challenges and opportunities of work in those spaces can inform such systems design research. We are particularly motivated to inform research into the design of systems that can aid DIYbio practitioners in achieving their diverse goals, but we also believe that our empirical research into community biolabs will provide a useful foundation for a range of possible HCI research questions.   

We seek to understand the conditions and constraints faced by those pursuing sustained work in community biolabs. 
In particular, we investigate: 
1.) What equipment and materials are available in community biolabs and how are these procured?
2.) What sort of  project work are people pursuing and what resources---labor or material---support project work? 
and 3.) What are the main challenges faced by those working in community biolabs?

To address our research questions, we interviewed 16 participants affiliated with 11 different community biolabs and supplemented these data with publicly available material from lab websites. We surfaced a number of themes grouped under `equipment, tools and materials', and `work'. 
We found that community biolabs are well-equipped for laboratory work, particularly molecular biology work, largely thanks to donations of equipment and materials from professional labs. 
Surprisingly, we found that time and effective means to coordinate group work are far more limiting than money or equipment. 
While some DIYbio practices involve people without professional training operating in low resource settings primarily using DIY equipment \cite{Tocchetti2014-bh}, that does not seem a fitting characterization for community biolabs. 
The people we spoke to were motivated by curiosity and a desire to learn as well as wanting to contribute their knowledge and skills to their immediate community of co-practitioners. 
Drawing on our findings, we discuss how community biolab members are creating new approaches to laboratory biology work, how laboratory equipment is procured and used in these spaces, and finally research opportunities for systems that support collaborative work in non-traditional settings for scientific practice.

\section{Background: Community Biolabs in context}

\begin{figure*}[htbp]
    \centering
    \includegraphics[width=\textwidth]{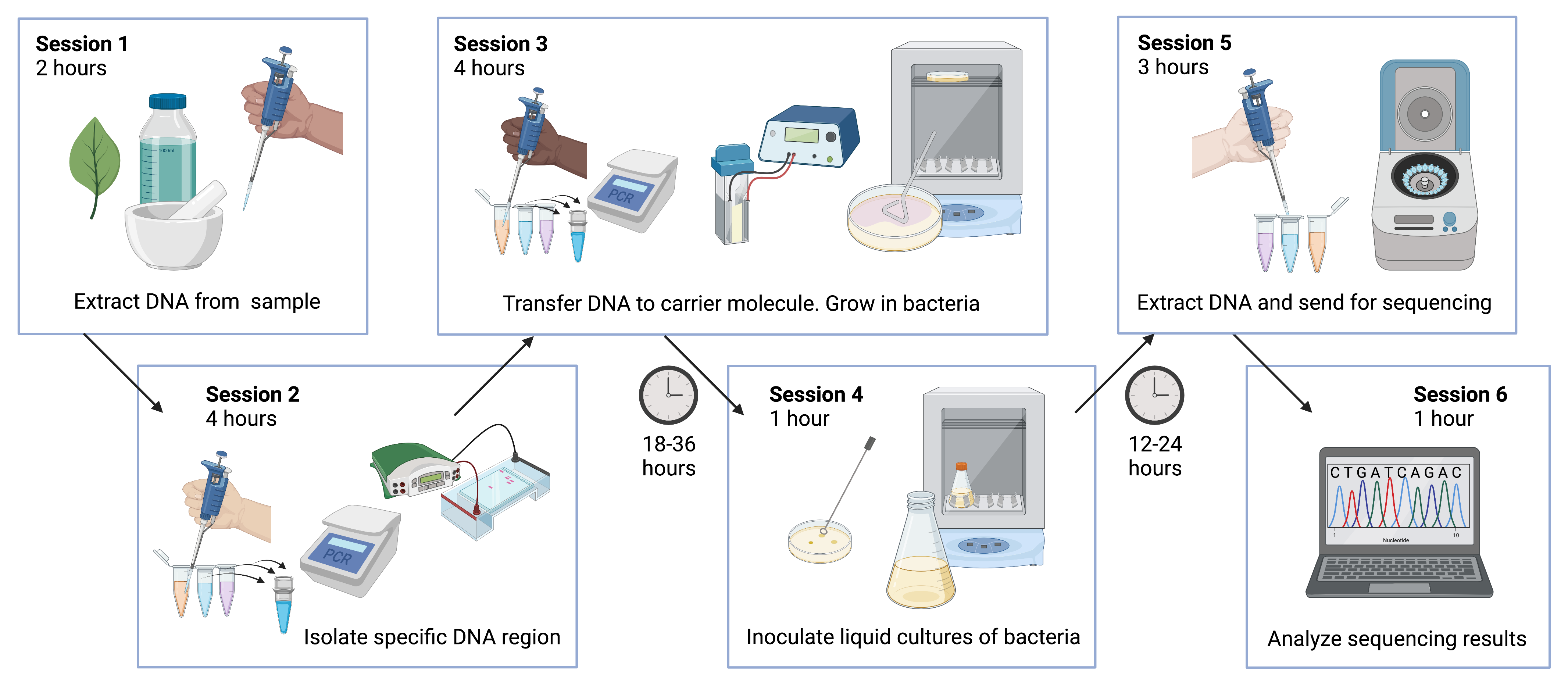}
    \caption{An example community biolab team project workflow. This workflow employs a variety of molecular biology techniques but is only a sample of the wide array of required equipment, instruments, and materials as well as the required knowledge and skills to make use of them effectively. To plan an effective workflow, the community member must rely on domain knowledge to choose appropriate timing, machine settings, chemical reaction mixes, etc. Furthermore, the steps must be performed sequentially because each step relies on the products of the previous step, and in some cases (clock icons) the time between work sessions must be kept within certain limits. Living organisms such as the bacterial cells used in this workflow tend to introduce time sensitivities into biolab workflows.  Data can be collected and analyzed at only a few steps. A failure could occur at any step but there are only limited opportunities to check for success since most processes of interest are impossible to directly observe.}
    \label{fig:workflow}
\end{figure*}

In the years following the launch of DIYbio.org in 2008, sociology of science researchers have engaged critically with the ethical, political and epistemological implications of DIYbio \cite{Keulartz2016, Delgado2013-xe, Kera2017,Wilbanks2017,doi:10.1177/0162243915589634}, while articles in the magazine sections of peer-reviewed life science journals were used to sound the alarm \cite{Wolinsky2009-vb} or sing the praises \cite{Alper2009-cj,Kean2011-bs} of the new approach to laboratory biology. A number of studies have sought to provide information about the on-the-ground state of the movement, for example by surveying its followers \cite{Meyer2020DIYbioHistory,Grushkin2009-ob} or the products of their work \cite{Landrain2013,Kera2012,Kera2017}, or through ethnographic studies of DIYbio groups \cite{Kera2011-db,Roosth2017-ev,Tocchetti2014-bh,Kaiying2016}.  The positionality of authors researching and writing about DIYbio varies with many  actively involved in DIYbio work or community organizing outside of their academic research, and some articles have been published with this explicit framing \cite{Scheifele2016-tk}.  From 2013-2018 BioCoder \cite{BioCoder} provided a platform for articles describing first hand reports of DIYbio work and related articles. 

In recent years, HCI researchers have begun to explore DIYbio, accepting its latent invitation to re-imagine the interaction space of human and non-human organisms with diverse machines and materials \cite{Asgarali-Hoffman2021-rc,Kuznetsov2015-ut,Kuznetsov2012-kt}. This has involved studying DIYbio practitioners operating outside of HCI \cite{Asgarali-Hoffman2021-rc} as well as integrating DIYbio into HCI practice \cite{Hamidi2021-wv,Fernando2016-tt}, and design research into systems that could support DIYbio work by better meeting the interests and contexts of practitioners \cite{Gome2019-gg,Fernando2019,Fernando2020-bq} .  

To the existing scholarship on DIYbio, we contribute findings on the contemporary practices and challenges of work in community biolabs. Our primary interest is to support DIYbio practitioners by informing relevant systems design research, though our findings are likely to be of interest to anyone with interests in the intersection of DIYbio and HCI. We provide here a brief overview of the technoscientific background for community biolab work as well as what is already known about these novel sites of practice.

\subsection{From molecular biology to the modern DIYbio movement }

Over the last 150 years DNA has increasingly moved to the center of the science of biology, transitioning from a mysterious acidic substance inside cells \cite{Dahm2005-sa} to not only the material out of which genes are constructed in nature, but also out of which new genetic programs might be rationally designed \cite{Nielsen2016-ik}. Within the last 50 years, a number of technological developments, including the Polymerase Chain Reaction (PCR) \cite{Bartlett2003-ia}, improved DNA sequencing, and practical, affordable DNA synthesis, have transformed laboratory biology. 
Molecular biology, in which DNA, RNA, and proteins are centered as laboratory materials and as a conceptual framework to explain life, is today a ubiquitous methodology within professional laboratory biology, whether in academia or industry \cite{Morange2000-io}.

The contemporary DIYbio movement was inspired by the ideas of 21st century synthetic biology \cite{Tocchetti2014-bh,Roosth2017-ev,Delgado2013-xe,Simons2021,Meyer2020DIYbioHistory} and was launched by individuals working in the Boston area, a center for synthetic biology research and industry \cite{Mac_Cowell2009-er}. Synthetic biology is a movement within professional laboratory biology, promoted initially by engineers and computer scientists who pursue the aspirational and problematized vision of rationally redesigning organisms, supported by the use of machine metaphors for genetic and cellular systems \cite{Roosth2017-ev}.  \citet{Roosth2017-ev} sees DIYbio drawing on the same tendencies of synthetic biology to pursue \textit{"intercalation of making and knowing"}, that is, to assert that building something is the same as knowing it. 
Indeed, we found that common workflows in community biolabs did incorporate standard molecular biology and synthetic biology techniques, as illustrated in the example workflow in Figure \ref{fig:workflow}.
The ongoing relationship between DIYbio and synthetic biology is also evidenced in the plenary lecture delivered by leading professional synthetic biology researcher Drew Endy at the 2020 Global Community Biosummit \cite{GCBS2020}, an important organizing event for DIYbio practitioners. Synthetic biology has been critiqued for promoting a ``\textit{molecular chauvinism}'' that promotes technological solutions and devalues the role of relationships among humans and to the non-human world \cite{Haraway2019Tools}. To some extent, following a different path, DIYbio brings the making of life into a domestic and personal perspective, and encourages a \textit{``crafty approach to biology''} centering the human practitioner as they work to achieve \textit{"epistemic knowledge ... generated through practical proficiency"} \cite{Roosth2017-ev}.
The contemporary DIYbio movement was imagined by its early practitioners and organizers as simultaneously fantastically expansive in its ambitions of engineering life and bringing biotechnology to everyone, as well as being small-scale and artisanal and occupied much more with the experience of working with organisms and biomaterials than with the technical outcomes of this work \cite{Roosth2017-ev}. 

In transferring things from the domain of the professional laboratory into new contexts, DIYbio practitioners are challenging technoscientific norms.
\citet{Cluck2015} found that DIYbio practitioners, as they create new spaces for their work, are \textit{``re-assembling what constitutes the means of production of scientific knowledge (a "laboratory")''}. 
As such, the sites of DIYbio practice, whether they be a private residence \cite{Tocchetti2014-bh}, a non-biowork-oriented hackerspace \cite{Kuznetsov2015-ut}, or an artist's studio \cite{Hamidi2021-wv}, are both laboratories of technoscientific work and, we suggest, the ``laboratories'' in which distinctly DIYbio ways of doing and knowing are explored. 
We believe that through studying these spaces and the work performed there, we gain insight into what could be possible for science outside of traditional institutions.

\subsection{What do we know about community biolabs}
In this study, we focus specifically on community biolabs among the sites of DIYbio practice. These are not the only relevant sites, others notably include home or garage labs \cite{Alper2009-cj, Wolinsky2009-vb}, as well as highschool classrooms \cite{Kafai2017}, design studios \cite{Kusnetsov2018}, and arguably online discussion forums \cite{Meyer2020DIYbioHistory}. 
"Community biolab'', while not a neologism of this paper, is the term we have chosen to define a diverse array of physical spaces and the organizations managing them. 
We offer a name and a working definition as a useful tool to navigate the complex reality of independent spaces managed by individuals with diverse practices and interests. 
\citet{Scheifele2016-tk} suggest that \textit{``community labs (biohacker spaces) [are] communal spaces that provide shared instrumentation, [...] reagents, management and communal projects for DIY-biologists.''} 
\citet{Wilbanks2017} refers to ``biohackerspaces'' as \textit{``communal spaces... for individuals to pursue scientific or biotechnological projects.''}
In this paper, we define community biolabs as dedicated spaces equipped for laboratory biology work, to which access is not restricted by qualification, that are independent of any traditional scientific institution, and not solely intended for commercial activity or structured educational offerings. 

By centering sites of practice in our interviews with practitioners, we contribute an exploration of the material resources of community biolabs (and how these relate to the work done in them) to a growing body of research on community biolabs. 
\citet{Landrain2013} explored the distribution of, and the practices taking place in, ``DIYbio labs'' almost a decade ago. 
More recently, \citet{MeyerBiohacking}, as part of a broader examination of ``biohacking'', lists and describes projects pursued at community biolabs. 
Detailed empirically grounded work on community biolabs has been provided by members and managers, such as the matter-of-fact recounting by \citet{Scheifele2016-tk} of the challenges involved in running a ``community lab'', and suggestions to others wishing to do the same.  
\citet{ScrogginsEducatingInLife} highlight the centrality of \emph{``acrimony and dissension''} as well as spotlighting the considerable work involved in the establishment of a new community biolab through the case study of BioCurious (www.biocurious.org). 
\citet{Asgarali-Hoffman2021-rc} have explored the nuanced relationship between bioartists and the organisms they work with, integrating the perspectives of community biolab organizers.
\citet{deLange2021} have suggested that community biolabs create unique and novel opportunities for the formation of relationships between amateur and professional scientists as well as between other stakeholders in order to respond to local community challenges. The scholarly work dealing with community biolabs over the last decade provides important context, particularly around the difficulties of establishing community biolabs and of project work. Our study contributes to this prior work by providing an updated view of the work taking place in community biolabs as well as taking a comparatively narrow focus on the material realities of work in community biolabs to better understand the constraints and challenges faced by lab members, and to surface new opportunities for HCI research. 

\subsection{Is work in community biolabs just "making" with biology?} 

The parallels between the Maker movement and the DIYbio movement are readily apparent, and the same can be said of makerspaces and community biolabs.   \citet{Delgado2013-xe} characterizes DIYbio based on a DIY hacker or maker culture: 
\textit{``[DIYbio is] very much about making and messing with things by combining wetware, software, and hardware in a low-cost fashion.''} Some community biolabs consciously identify themselves as biology makerspaces or as biohackspaces. Community biolabs are at times presented as a subset of makerspaces and DIYbio as a sub-movement of the Maker movement \cite{Menichinelli2019-wx}. 
However, DIYbio has also been framed as primarily a movement within natural science practice. \citet{Keulartz2016} suggest that \textit{``DIY-Bio does not represent new science but a new way of doing science.''}. DIYbio has been described by practitioners as an alternative to institutionalized science, despite at the same time in many ways acting to support these norms \cite{Simons2021}. \citet{Golinelli2016} offer that DIYbio can be understood as \textit{``an engagement with advanced life sciences outside of the traditional professional environments of the university-industrial complex.''}. \citet{Kera2012} offers a hybrid framing, describing grassroots R\&D labs that have \textit{``revived the bottom-up relation between community building and experimenting with new knowledge and technologies''}, encompassing sites of DIYbio alongside maker, hacker, and art and design practices. 
For this work, we approached community biolabs assuming that we would find a mixture of thing-making and knowledge-making, and in our interviews we queried both the extent of maker practices within community biolabs as well as the unique features and requirements of biowork. 

\subsection{Providing an empirical foundation for design research to support work in community biolabs} 

We are interested in supporting DIYbio practitioners by providing context about the material realities of DIYbio work to inform future systems design research. 
Among the concerns of context-aware designers found by \citet{Bauer2014} are the intentions, preferences, and prior knowledge of users, as well as considering the physical environment and the way that the designed system will exist within that space and communicate with users. 
Novel contexts can profoundly alter both how the work is performed as well as its significance. \citet{Dourish2004} points out in a consideration of the role of context in systems design that \textit{“the word ‘hello’ can be not only a greeting, but an inquiry, a rebuff, a joke, an exclamation, etc., depending on how, when and by whom it is used.”} 
Analogously, smearing a petri dish of agar jelly with bacteria and letting the colonies grow in an incubator could be an artwork, an experiment, or a student project depending on the context. \citet{Kuznetsov2012-kt} and colleagues have previously demonstrated examples of how the context in which biowork is performed and presented can radically reshape its meaning and suggest alternative goals for laboratory biology systems designers. 
Therefore, we believe that surface level notions of what work is performed in community biolabs and under what conditions, i.e., the sort of data available by merely browsing biolab websites, is not a good basis to inform future engagement between HCI and DIYbio work at community biolabs.  We contribute our findings and discussion from interviews with contemporary users and managers of community biolabs, adding to the small but growing body of HCI research into the diverse, technology-rich ways of working with biological materials that is encompassed within the term DIYbio. 

\section{Methods}
In this section, we provide detail on our positionality, data collection, and data analysis methods
\subsection{Positionality}

To provide context to our inquiry and analysis, we are including details on author background and positions \cite{10.1145/3025453.3025766, 10.1145/1978942.1979041, 10.1145/3443686}.
The first author has extensive experience both as a formally trained professional laboratory biologist and as community biolab member and manager. (\SoundBio; Orlando de Lange). 
We leveraged his position to recruit participants and his knowledge to provide context for understanding the specific lab materials, techniques, and equipment.

\subsection{Method selection}
We are interested in what people do, why they do it, and the context within which they work. We felt that a semi-structured interview method, compared to a survey or structured interview, would provide the appropriate amount and quality of data since it allows for interviewer follow-up questions and for the interviewee to explain their answers at length. As one of the authors has extensive experience working in professional and community biolabs, much of the valuable context that would be provided through observational methods was not necessary to address the goals of this study, and rather we considered it more important to prioritize collecting data from a greater number of participants working at a greater range of community biolabs, in terms of geographical location and operational model. Our methods were vetted by our institute's IRB.

\subsection{Identifying community biolabs and recruiting participants}

\begin{table*}[htbp]
    \begin{tabular}{l l l l}
 \hline
\textbf{Lab} & \textbf{Participant Name} & \textbf{Participant Role in Lab} & \textbf{Location} \\ 
 \hline
 Biotech Without Borders & Ibrahim Dulijan & User & New York, NY, USA \\ 

 Boslab & Kellen Andrilenas & User, Manager, Educator & Boston, MA, USA \\

 Boslab & Wendy Pouliot & Manager & Boston, MA, USA \\

 BUGSS & Anonymous & User & Baltimore, MD, USA \\

 BUGSS & Lisa Scheifele & Manager, Educator & Baltimore, MD, USA \\

 BUGSS & Yann Huon de Kermadec & User & Baltimore, MD, USA \\

 Cambridge Biomakespace & Anonymous & Manager, Educator & Cambridge, UK \\

 Counter Culture Labs & Tim Dobbs & User, Manager, Educator & Oakland, CA, USA \\

 Hive Biolab & Prince Edem Samoh & User, Manager, Educator & Kumasi, Ghana \\
  
 London Biohackspace & James Phillips & User, Manager, Educator & London, UK \\

 Open Science Network & Scott Pownall & Manager & Vancouver, Canada \\

 ReaGent & Anonymous & Manager & Ghent, Belgium \\
 
 SoundBio Lab & Anonymous & User, Manager,  Educator &  Seattle, WA, USA \\

 SoundBio Lab & Chris Gloeckner & User, Manager & Seattle, WA, USA \\

 SoundBio Lab & Garima Thakur & Manager & Seattle, WA, USA \\
 
 TriDIYBio & Tom Randall & Manager & Durham, NC, USA \\
 \hline 
    \end{tabular}
    \caption{Lab and Participant information}
    \label{tab:participants}
\end{table*}

For this study, we define a community biolab as a physical lab space equipped for life science work that is run as a not-for-profit enterprise independent of any large institution such as a university or biotech company. 
In addition, access must not be limited by professional qualifications and the space should not be primarily oriented towards start-up business support, i.e. not primarily a biotech incubator. We reemphasize here that there is no widely adopted external or community-internal definition or model for what constitutes a community biolab.  Even the labs we included under our narrow criteria represent a range of different models, informed by different goals and worldviews, though, as we found, reasonably unified in terms of equipment and regular activities. 
de Lange has four years of experience managing community biolabs and regularly attends community gatherings. 
We used his prior knowledge to establish a shortlist of relevant lab spaces, and supplemented this with a broad keyword internet search. 
We identified a total of nineteen labs worldwide that we could confidently confirm as in scope and still operational. Additional labs were identified that, based on online presence or personal correspondence, seemed to be no longer operating or otherwise did not meet our inclusion criteria. We are confident that we were able to identify all or almost all in-scope labs, though there is reason to suspect that additional in-scope labs exist in regions other than Europe and Anglophone countries.

We reached out to people at the labs we knew to be in scope as well as others that we were unsure of. We asked to interview those who manage lab spaces and are in a position to speak broadly about the work pursued in their lab, as well as people who have been engaged in sustained project work that uses the lab space and its material resources.  
We used existing relationships or contacted potential participants via publicly available email addresses or social media accounts. 
We present data from interviews with sixteen people working at eleven different community biolabs in North America, Europe, and Africa, detailed in Table \ref{tab:participants}. These include nine of the nineteen labs we had already defined as in-scope, as well as two additional labs that on careful evaluation fit our criteria closely enough to warrant inclusion. We believe that our sample includes data from members of roughly half of the extant community biolabs that we were able to identify. 
Twelve participants identified as White or Caucasian, two as Asian, one as Black, and one preferred not to identify. Eleven participants identified as men, five as women. 
All participants had undergraduate or graduate level STEM education, and fourteen of them were employed in a STEM field at the time of the interview.
We concluded that one of the labs from our initial recruitment, based in Los Angeles, once described by an interviewee, fell outside the scope of this study because it primarily served as a space for the work of the interviewee's private company, despite sharing many features of other labs we included and being managed by an individual with extensive experience using and managing other community biolabs in prior years. This interview and lab were dropped from our sample, and are not included in the final dataset of sixteen interviews from eleven labs. This example is reflective of the diversity of sites of DIYbio practice, and illustrates that the definition of community biolab we have used is just one possible definition, and that further empirical work into the diverse forms of DIYbio practice is needed.

\subsection{Data collection and analysis}
We asked participants to complete a short intake questionnaire providing information on demographics, prior expertise or education, and their role (or roles) in relationship to their current community biolab (manager, user, educator).
Then we conducted 45 minute semi-structured individual interviews of all sixteen participants with one or more authors on Zoom. 
We used different pre-planned interview protocols for when a participant considered themselves primarily a lab manager or a lab user, while acknowledging that such a clean distinction was not possible to make for most participants, allowing for ad hoc elaboration on questions. Participants who identified as educators also always held the roles of lab user or manager, so a separate interview protocol was not necessary for educator roles.
We questioned lab managers about the space, the equipment within it, and the work involved in managing the space. 
We asked lab users to describe in detail the work they conduct within the lab space, the resources they use, and the challenges they face. 
In all cases participants were asked about the motivations for their work in the community biolab. 
In addition to interview data, we also collected photos and other documentation (project websites, albums, etc) that participants voluntarily shared.
Finally, we cross-referenced interview data with photos and information from the publicly posted websites, blogs, and social media accounts belonging to the respective labs.

We transcribed the interview recordings automatically using temi.com, and then inspected and manually corrected each transcript for accuracy. Following a thematic analysis method \cite{BraunVirginia2006Utai, BraunVirginia2019Rort,SaldanaJohnny2021Tcmf}, we developed a codebook of 33 provisional codes in collaborative sessions, which expanded to 44 codes and subcodes during inductive coding of transcripts. Our codes captured several types of data, including descriptive (physical equipment, software, costs, space management, materials, expertise...), process (help, learning, time management, external collaboration, automation, coordinating work...), and conceptual (barriers, vision/worldview, unrewarding labor, existential threat, motivations...). Each transcript was coded by one of three authors using Taguette software, and then separately reviewed and re-coded by the remaining two authors for intercoder agreement and reliability. During secondary analysis discussions and writing, we themed all of the coded data into two categories (Equipment/Tools/Materials, and Work), with prominent subthemes of each category supported by frequency counts of applicable codes and relevance to our research goals. These themes and subthemes are presented in our Findings, and our interpretations and suggestions are presented in the Discussion.

\subsection{Limitations of our methods}

\subsubsection*{COVID-19 pandemic.} Interviews were conducted from Spring 2021 to January 2022, meaning that community biolabs had been managing COVID-19 pandemic related restrictions for over a year in most cases. We asked about the impacts of the pandemic in all cases. Participants reported severe limitations on access to the lab and group activities such as workshops, but some project work had been able to continue. If work had ceased during the pandemic, we asked interviewees to report on the pre-pandemic situation. 

\subsubsection*{Geographical limitation.} 
The majority of the organizations we identified as in scope are based in the USA and this is reflected in the locations of our interviewees. We are aware of groups in India, Africa and Latin America that we could not unequivocally determine to be in scope based on publicly available information.  The use of prior knowledge and  English language keyword internet searches may have led to us missing some organizations entirely. We feel that our results are likely to provide a representative sample of conditions in community biolabs in 2021, but considering that even within our sample we found a great deal of diversity in organizational models, practices, and perspectives, caution should be taken in extrapolating any particular finding to labs not surveyed or into the future or past of community biolabs.

\section{Findings}

We present findings on the nature and context of technoscientific work in community biolabs. We pay particular attention to issues of relevance to systems design and human-computer interaction research. 

The community biolabs described by participants ranged in size from a single 500 square-foot room to a space ten times that, spread across multiple separate rooms. Images of some of these lab spaces are shown in Figure \ref{fig:labs}. The community biolab spaces were arranged following the norms of professional life science laboratories, featuring benches for project work and specialist equipment on hand, sometimes grouped into dedicated work areas, e.g., media prep area. Chemicals were stored separately from other materials.

From participants identifying as lab managers we learned that significant organizational work, often provided by volunteers, is required to keep community biolabs operational. In some cases, one or two people were responsible for day-to-day supervision and support of lab operations, as well as administrative tasks such as updating the website or filing taxes. In most cases, lab management was entirely a voluntary activity. Four of eleven labs reported having one person in a full or part-time paid position supporting project work, three of which were general lab managers and one was supported by a grant to work on a specific project. Estimates of total lab management time investments were reported as fluctuating week to week, from as little as zero for a quiet week but in most cases at least 10 hours a week. We found that some lab managers feared for the long term sustainability of their organizations due to threats of financial insecurity or volunteer labor burn out. 

Structured educational activities were mentioned by most participants but we purposefully focused our interview questions on project work. Community biolabs organize workshops, courses, talks, and schools' outreach activities, and the distinction we have drawn between education work and project work is not necessarily easy to draw in practice when considering the complete set of activities supported by any given community biolab.  

With a single exception, participants told us that the community biolab offered them resources they would have no other way to access in their geographic area. 
Key affordances of these shared sites of practice include equipment, space, opportunities to procure otherwise restricted materials, relevant provisions for safe work, and opportunities to learn with and from co-practitioners, as well as providing some users with a valuable sense of contributing or giving back. 
All labs provided safety oversight, as explored below, while some provided safety training to all new members. One participant with a background in life sciences valued being in a "lab setting" and \emph{"the whole embodied knowledge that comes back as you feel the place"}. These sites, though few in number and limited in space, provide considerable meaning and value to those who use them.

\subsection{How do community biolabs use and manage equipment, tools, and materials}

\begin{figure*}[htbp]
\centering
{\includegraphics[width=1.0\linewidth]{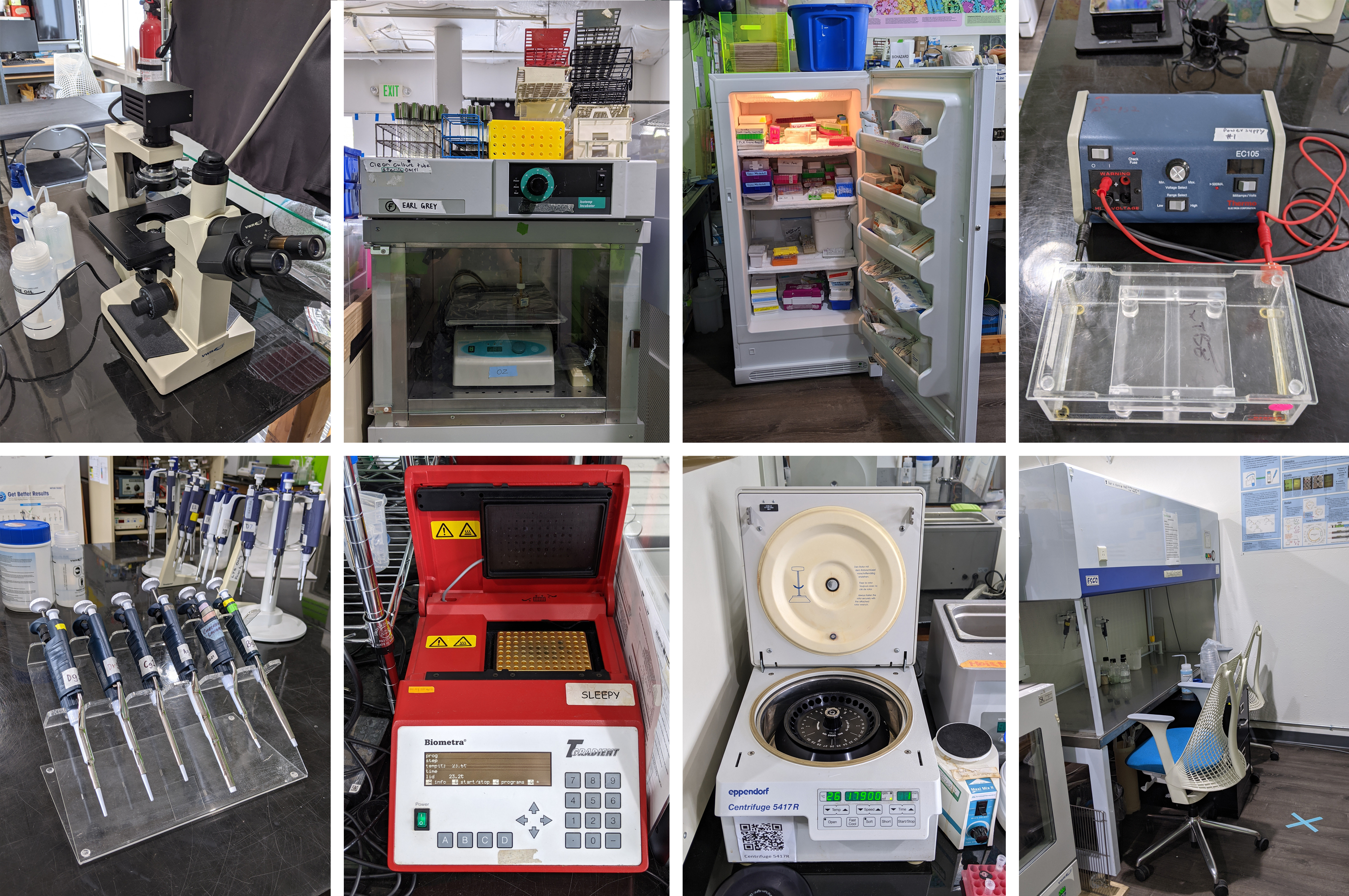}}
\caption {Some of th especialized equipment commonly found in many of the community biolabs in our sample. Nearly all of our labs acquired professional-grade equipment through donations or secondhand purchases. These photos were all taken at \SoundBio. Top row, left to right: Light Microscope; Incubator with orbital shaker positioned inside; Fridge; Gel electrophoresis equipment. Bottom row, left to right: Set of micropipettes; Thermocycler; Benchtop microcentrifuge; Laminar flow hood.}
\Description[Eight examples of equipment commonly found in a community biolab.]{Photos include: a microscope on a table, an cube shaped incubator with an orbital shaker inside it, a standard size refrigerator, equipment for gel electrophoresis on a tabletop, a laminar flow hood that takes up the corner of a room, a benchtop microcentrifuge with the lid open, a small thermocycler, and a collection of handheld pipette tools.}
\label{fig:equipment}
\end{figure*}
%

Physical tools (esp. those controlled via a digital interface) are a productive site of intersection between DIYbio and HCI \cite{Kuznetsov2015-ut, Fernando2020-bq, Gome2019-gg}. We therefore asked participants to list and describe the tools they work with, what they use them for, and how they were obtained. 

The exact set of equipment and materials varied between labs, as did the range of projects and other activities occurring in the spaces. However, a unifying feature is that all of these spaces have been equipped to support the broad domain of laboratory life science practices known as molecular biology, encompassing work that centers on the manipulation of DNA, RNA and proteins, as well as culturing microbes and other common lab organisms. 
Figure \ref{fig:equipment} shows some of the equipment that was listed by most study participants. Some community biolabs included benches with built in controlled airflow to allow for sterile work or work with potentially biohazardous materials such as animal cell cultures. Community biolabs provide dedicated space and professional-standard equipment for laboratory biology work.

\subsubsection{Materials are controlled to ensure safety}

All but one participant identifying as a lab manager shared with us that lab safety is a central concern of their role. Every lab had safety systems in place, some formal (e.g., required safety training) and some informal (lab manager advises on and prohibits certain work as needed). A minority of lab managers expressed anxiety around the possibility of a safety incident due to a random accident. 

Work in community biolabs is kept safe either through the adoption of rules and standards shared with members, or by the bespoke review and intervention of expert lab managers. We asked all participants about the importance of safety and about the specific mechanisms they use to prevent unsafe work. Participants from five of eleven labs mentioned voluntary adherence to the Biosafety Level 1 standards of the National Institutes of Health, or European equivalents. Five described to us that they restrict the materials that can be brought into the lab space, and three described restricting specific lab activities. In three cases, participants shared examples of reviewing and rejecting project proposals on safety grounds, and three labs mentioned a designated safety director or committee. Three of the participants in our study described providing individual oversight and targeted safety intervention to every project in the lab they manage, tailoring the degree of oversight to the perceived trustworthiness of the lab member. We found that while safety is a near universal priority for community biolabs in our study, the mechanisms for keeping work safe vary from lab to lab, though a dominant theme was early or preemptive intervention to stop work with specific materials perceived to be unsafe. 

\begin{quote}
    \Lisa: We actually have a white list of what we allow in the lab. So there are certain things that, you know, salt, you can bring that in. No problem. Um, and then anything that's not on the white list, you've got to tell us, and you've got to get checked over with us. And for a new member when they joined, we have required them to kind of write out a paragraph, what's your project. So we understand it.
\end{quote}

Machines and tools are the dominant concern for makerspace safety systems, contrasting with the emphasis on materials (chemicals and organisms) in biolabs, a difference made explicit in a 2016 proposal to adapt biolab safety norms to makerspaces by swapping out controlled materials for controlled equipment \cite{Klein2016}. As an example of how materials are a locus of safety control, several lab managers described examples where a project had been disallowed since it required work with mammalian cell cultures which can become reservoirs for pathogenic microbes or viruses that can then pass to humans. Chemicals can present a range of immediate safety hazards during use, but also must be stored and disposed of safely, creating not just anxiety but unwanted labor for lab managers. Concerns were also raised in relation to a few types of equipment such as high-speed centrifuges and autoclaves (pressurized steam-based sterilizers).

\subsubsection{The cost of commercial lab equipment is not a major constraint}

Every lab we heard from sourced equipment second-hand through donations, comprising the vast majority of equipment in all but one of the labs within our sample. Local biotech companies and university labs were cited as sources of equipment, and at six of the labs in our sample equipment was available that had been purchased, loaned, or donated directly by lab members. 
One lab manager had their own well-equipped home lab and described ferrying some equipment and materials back and forth. 
Equipment had also been purchased second-hand at most of the labs in our sample, with eBay repeatedly cited as a platform for these activities. The only lab in the developing world within our sample was also the only example of a lab equipped largely with newly purchased equipment, sourced via a major collaboration with a partner in another country. 

Rather than struggling to obtain expensive professional-standard lab equipment, as we might have expected, in many cases participants lamented the overabundance of equipment and materials, because it was leading to space constraints. 

\begin{quote}
    \Wendy: \textit{We have personally reorganized the lab about 14 times because it's trying to fit it in a very small space, uh, and manage storage...And now we've taken in so many donations during COVID that it's scattered in everyone's storage units...It's a hot mess.}
\end{quote}

Eleven participants spoke of equipment in the lab that was broken, unreliable or had been repaired. We did not collect data that would allow us to form a clear idea of how community biolabs are coping with the burden of equipment maintenance, which may represent more of a problem as time progresses and the still relatively new, second-hand equipment in community biolabs ages.

\begin{quote}
    \Andy: \textit{The equipment itself is mostly off of eBay or is donated by people who don't want it anymore and take it as a tax write off. So it's old equipment, some of it's iffy, some of it stopped working, uh, and you know, some of it started working again when you hit it right. Some of it didn't. So, you know, there were always challenges like that.}
\end{quote}

Obtaining lab equipment of sufficient variety, quantity, and quality to allow members to pursue technoscientific work that they find meaningful does not seem to be a challenge for community biolabs in our study. However, participants cited multiple examples of specific types of desired equipment that they struggled to obtain. These included specialist freezers able to store samples at -80°C (required for long term storage of microorganisms); more powerful centrifuges; and ventilated workbenches either to trap and export fumes (fume hood), create a sterile work area (laminar flow hood), or to block the escape of dangerous organisms or biomaterials from the sterile work area (biosafety cabinet). Two participants wanted to obtain a machine for high-throughput DNA sequencing. 
None of the labs in our study had machines for DNA sequencing; we learned about work at three labs that involved sending off DNA samples to be sequenced via external collaborators or services. 

Specialist equipment is not the only thing crucial to work in community biolabs. Project work also requires labware (e.g., glass beakers), a range of specialty disposable plastics known as consumables (e.g., micro-pipette tips and small resealable plastic tubes), chemicals and reagents, and the organisms and biomaterials that are the primary focus of any given project. Some of these items are depicted in Figure \ref{fig:workflow}.  We focused our inquiry on equipment, but three lab managers shared that they also received substantial donations of consumables and chemicals. Specialist reagents and chemicals were each mentioned as cost barriers and one participant talked about the difficulty they encountered convincing some companies to ship these specialist items to a non-university, non-industry lab. 

We found that community biolabs were successfully sourcing an abundance of high-end laboratory equipment for no- or low-cost. However, lab members in many cases do still face barriers in accessing the equipment and materials they need, in functional condition, and at a price they can afford.

\subsubsection{Making equipment creates new opportunities for meaningful work}

Making is common in community biolabs but DIY equipment or tools were rarely described as crucial to support scientific project work. When asked about making practices, 14 participants were able to cite a total of 35 specific examples of research tools or equipment that they or others had made in-house. 
Examples included simple re-purposing, such as using a reptile terrarium as a laboratory incubator, or an aquarium as a makeshift sterile work bench, or fabricating a camera mount for time-lapse photography. More sophisticated examples included a DIY shaker-incubator (for cultivating microorganisms) and a DIY gene gun (for delivering lab-prepped DNA into cells). In another community biolab, a group of members had refashioned a 3-D printer into a working bioprinter with the intention of encapsulating microalgae in a gel matrix. This not only created new possibilities for project work, but attracted significant interest from other lab members, acting as a nucleation point for new social connections. 

Making lab equipment is rarely driven by the high cost or non-availability of equipment, rather the process itself was attractive to participants as a fun and enriching activity. Creating frugal lab equipment is seen as a compelling challenge by some community biolab members.

\begin{quote}
    \David: \textit{During that week I was fascinated by it [3D printed centrifuge]. But then after that I kind of just stopped using it just because I didn't really have a need for it. It was just for fun and trying to figure out how this works.}
    \end{quote}

Six of the labs we heard from were housed within or in the same building as a makerspace, and a collaboration with a local makerspace was mentioned by a participant from one of the other labs. In following up on the nature of the interactions it seemed that the biolab activities were almost totally independent from the makerspaces and most biolab members weren't transitioning back and forth. Nevertheless, the availability of equipment, materials or expertise from the makerspace and its users was accessed at least intermittently to support maker activities in the biolab, according to our interviewees:

\begin{quote}
    \Wendy: \textit{We have a bioprinter that was hacked from a 3-D printer and that was made by a group [...] who crafted that in collaboration with another makerspace so that we could obtain the necessary parts since we don't [...] have a machine shop or any capability of really making some of the components.}
\end{quote}

While it seems that most of the equipment used in community biolabs has been sourced ultimately from a commercial manufacturer, that does not mean that DIY lab equipment is an irrelevant concept as value can be derived not only from the product but also from the process itself. Relationships between community biolabs and makerspaces are already common and productive, if limited in scope and depth.

\subsubsection{Enthusiasm outstrips applications for automated liquid handling} 

Machine automation is used regularly in community biolabs, for example in the form of thermocyclers (see Figure \ref{fig:equipment}) that automate the rapid cycles of heating and cooling required for a PCR reaction. However we were particularly interested in whether robotic automation would be prevalent in community biolabs, in line with growing adoption within professional biolabs, particularly in industry and clinical settings and to much lesser extent in academic research labs \cite{Holland2020}.  In fact a majority (seven) of the labs we heard from listed an automated liquid handling platform, most commonly an OpenTrons robot, among the equipment available in their lab. In addition, many interviewees expressed interest in, and broad support for, robotic automation though few had used robotic automation or had any concrete plans to do so in the future. Desirable affordances of automation offered by participants included common refrains around automation in professional biolabs \cite{Holland2020} such as speed, efficiency, scale, safety, reproducibility, accuracy. Additional affordances hypothesized or realized by participants included skills development for lab users, collaboration opportunities, facilitating bioart work, and being able to work remotely (from the lab) or at a safe distance from harmful materials. 

\begin{quote}
    \Kellen: Being able to select colonies as candidates and do some of that handling would ... allow an individual to be able to do it at scale, as opposed to now, like, even, even just trying to get a 96 well plate of plasmids cloned with a team of sort of untrained people. It was a challenge and we didn't even finish it all the way because people were like, cool, I'm done now. And they're like, but wait, there's more, wait come back, I have more labor. So having a robot do that would be like aces.
\end{quote}

\begin{figure*}[htbp]
\centering
{\includegraphics[width=1.0\linewidth]{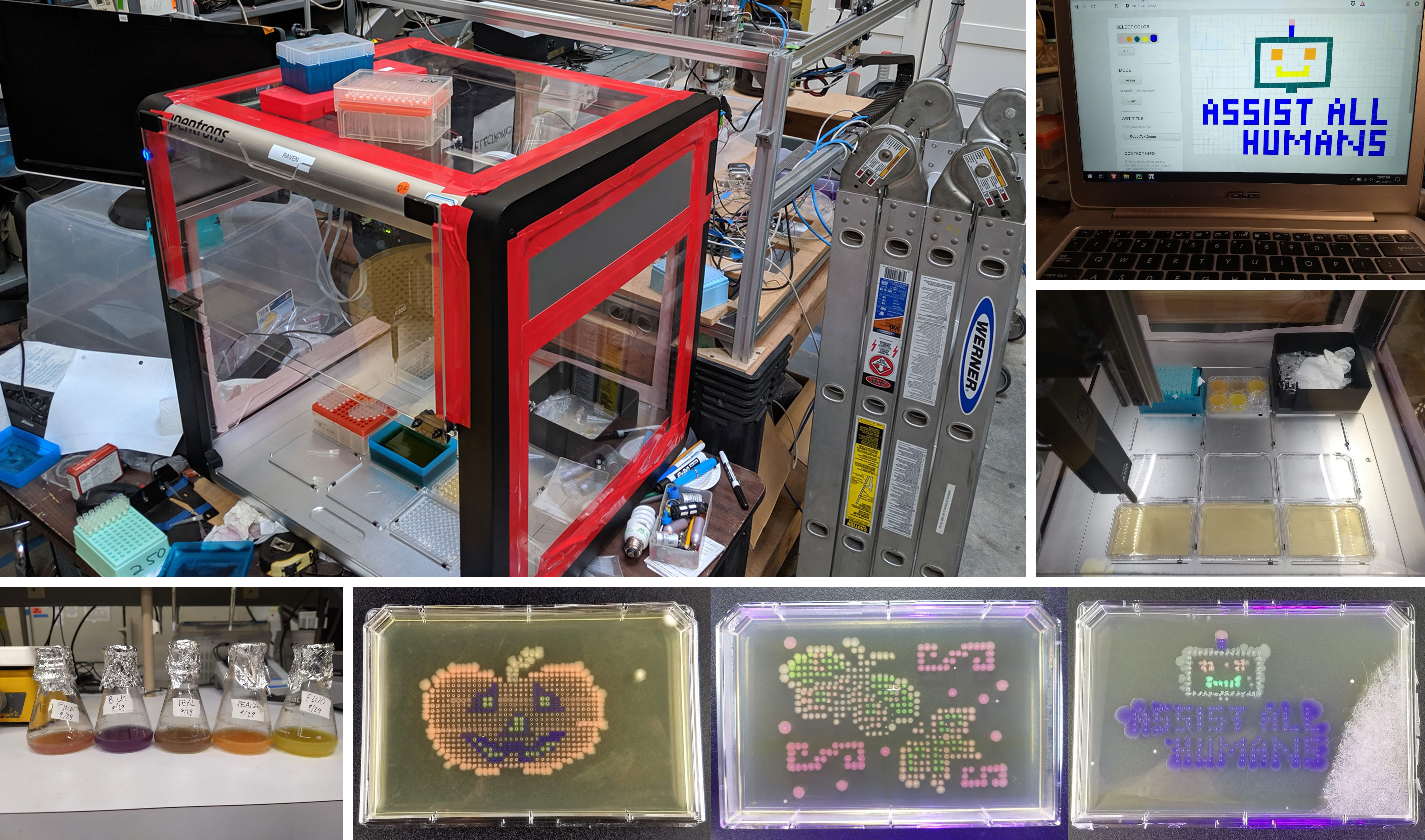}}
\caption {The bioartbot.org project at \CCL was the project with the most extensive use of automation in our study. Participants in the project design an image via a website, which is then printed in bacteria using the lab's OpenTrons OT-2 liquid handling robot. Clockwise from top left: the Opentrons OT-2; a design on the bioartbot.org interface; pipetting bacteria onto a gel substrate using the Opentrons; art pieces after incubation; colored bacteria prepared in flasks.}
\Description[Process photos of the bioartbot.org project.]{The Opentrons machine is cube shaped with transparent sides taped on, sitting on a cluttered tabletop next to a computer and a ladder. The photo of the website interface shows an example drawing of a robot face that says "assist all humans." An action shot of the Opentrons automatically pipetting bacteria into rectangular plates containing a gel substrate. A row of glass flasks show different colors of bacteria in solutions. Three finished bio artworks are colorful drawings of a pumpkin, some butterflies, and the Assist All Humans robot face.}
\label{fig:bioartbot}
\end{figure*}

One interviewee at \CCL was making extensive use of the OpenTrons OT-2 robot as a platform for a project aimed at engaging non-lab members through a website to design an image to be printed in bacteria onto a solid growth medium by the lab robot (\href{http://www.bioartbot.org/}{bioartbot.org}), as seen in Figure \ref{fig:bioartbot}. Interestingly, this participant’s engineering background and expertise was such that the robot offered an attractive entry point to work in the community biolab. The project also showcases the potential for automated machines to facilitate the precision execution of creative expression and to be a platform around which social interactions can be organized. In two other labs the liquid handling robot was being lightly used with a view to heavier use in the future. 

In some cases, participants suggested that automated liquid handling was unlikely to be useful in a community biolab setting when weighed against the drawbacks. The difficulties in setting up, using, or maintaining the machine were mentioned by three participants, while two cited a lack of available bench space. Several participants also mentioned that, despite enthusiasm, there isn't currently a need to work at the scale afforded by a liquid handling robot. 

\begin{quote}
    \Abigail: I think the reason we don't have automation on a widespread call for it in our labs at the moment ... I guess I'm thinking of the OpenTrons stuff here ... I think we have so many one-offs that it hasn't rewarded, um, making something that's highly repeatable, you know, being taken out of, out of our hands.
\end{quote}

Robotic automation currently occupies an ambiguous position within community biolab work. Systems are available and lab members are interested in using them, but actual usage seems to be limited.

\subsection{What work is performed in community biolabs, and what are the conditions and motivations that surround it}
We are interested in the activities taking place in community biolabs as part of ongoing project work pursued by lab members. Participants talked about projects with a diverse range of goals, but day-to-day activities often involved the manipulation of microorganisms and DNA. Examples of techniques employed by lab users included culturing  microbes, mostly yeast and bacteria, followed by precision purification of DNA or other biomolecules, such as insulin, from microbial cells. Several participants described using gel electrophoresis to separate DNA molecules within a physical gel substrate (as depicted in Figure \ref{fig:gels}) and others described preparing DNA samples to be sent to collaborators or companies for DNA sequencing. Stitching DNA fragments together to make circular plasmid DNA and then introducing that DNA into bacterial cells---a process called genetic transformation---was also mentioned. Many of the tools needed for these tasks are shown in Figures \ref{fig:workflow} and \ref{fig:equipment}.

Some of the projects described to us involved clearly defined goals, with structure such as check-in meetings to assign near-term priorities and provide accountability, though participants also mentioned numerous examples of projects that dramatically  shifted goals or activities unexpectedly, or examples of work abruptly ceasing.
Project work described by participants included both individually conceived and executed projects as well as cooperative projects with a range of team sizes.  
We did not find that cooperative project work involved clearly defined roles or hierarchies resembling the traditional principal investigator model of university life science research labs. 

One participant spoke at length about the value of a horizontal team structure and their desire to avoid hierarchies, despite the logistical challenges this created. The paths and social structures of projects at community biolabs must respond to the realities of these organizations with new members coming and going, changing interests and availability and a lack of clear, hierarchical roles.  Participants emphasized the importance of learned skills ranging from laboratory techniques and use of specific equipment through to failure tolerance, patience, and scientific thinking.  These expertise barriers, together with the time sensitivities of biowork (long workflows involving many steps, long incubation times, and certain steps to be performed within certain time windows such as shown in the example workflow of Figure \ref{fig:workflow}) were raised as significant complications for the pursuit of cooperative work, particularly work between individuals with different prior training and experience levels.  

\subsubsection{Community biolab project work is time sensitive and time consuming}

One participant was in the rare position of having been awarded a grant that allowed him to work full time on his project. Some users are retirees with more free time to devote. But in most cases, users of community biolabs were working in their free time alongside jobs, studies, child-care duties or other competing demands. This was exacerbated by the additional time required to get to and from the lab space, particularly for labs sited in large cities, for each lab work session. This can lead to relatively large time investments in travel time to accomplish quick, routine, tasks. 

\begin{quote}
    \Lisa: \textit{I think, you know this about community labs, people come from far away sometimes because there's only one in a metro area usually, so people have to travel a bit of a distance and [...] with biology, there's stuff that has to get done every day. And sometimes it's literally five minutes of work. Like you've got to scrape a bacterial cell, put it in some broth.}
\end{quote}

Many common techniques used in biolabs also require long periods of waiting. For example, a PCR reaction, depending on the precise thermal cycling programming used, may take 1-3 hours to complete. Laboratory techniques often involve chemically complex mixtures of materials, undergoing reactions occurring unseen and for which success or failure can only be assessed at the end of the process via a separate assay. 

\begin{quote}
    \David: \textit{If you're just trying to PCR a really long fragment, it can take a couple of hours, and chemical transformations take a lot [of time] too and they don't always work, you have to redo them sometimes. And so that gets pretty frustrating, and just eats up a lot of time. Yeah, biology is just really finicky.}
\end{quote}

Lab users working with living organisms had to attend to the sensitivities of growth rates, and the fact that in most cases organisms can only be stored for limited time periods before dying due to lack of nutrients or otherwise becoming unusable. Figure \ref{fig:workflow} shows a hypothetical workflow similar to several examples shared during interviews, and illustrates the way that lab work with laboratory organisms requires practitioners to devote not only significant time but also to carefully structure that invested time. 

\subsubsection{Expertise requirements complicate cooperative work}

Most of the participants in this study had completed a graduate degree in a STEM topic or were currently employed professionally in a STEM field, in accord with a 2013 survey of the broader DIYbio community \cite{Millet2013-ff}. For some participants, the community biolab was a unique space that allowed them to retain and use their professionally developed expertise. For other participants, the community biolab was a space to develop new expertise, either through project work or inspired by encounters with other community biolab users with different areas of expertise.

\begin{quote}
    \Abigail: \textit{I met some bioinformaticians in the space and now I am like a software engineer/bioinformatician, and I would never have realized that that was a career option without the space.}
\end{quote}

However, interviewees also cited times that the lack of specific technical knowledge or skills created a barrier to progress on a project. The diverse laboratory biology practices being pursued in community biolabs mean that no one individual, even with a professional laboratory biology background, is likely to be able to provide advice or training to all lab users. 

\begin{quote}
    \Kellen: \textit{I have PhD in cell and molecular biology. So this is what I do. It, I think for other folks in the lab, it is definitely a stumbling block ... cause they're like, oh, I want to do this thing, and we're like, that's great, now we get to teach you how to do that thing, which is a little more challenging. There's certainly areas where I do not have sufficient knowledge, like my knowledge of metabolic engineering...that's not my area.}
\end{quote}

The problem of generating and sharing expertise in community biolabs is vexing. Essential to the model of these spaces is that no professional qualifications are required to become a member, but this does not mean that members do not appreciate the importance of expertise and the difficulty for new members in acquiring the required expertise while wanting to immediately make progress on their projects. Nine participants in the study talked about the importance of sharing expertise or training others in order to allow project work to advance. A significant barrier exists for those even just wishing to participate in work within a project designed by another person; any given project in a community biolab is likely to require expertise in using multiple machines, instruments, materials and reagents as well as having enough knowledge about the underlying biology and how the methods work to be able to troubleshoot and interpret data. To be able to be the primary designer of a project was presented as an even greater hurdle, and one that might seem deceptively simple to those with no laboratory biology experience. The time it takes for members to gain and share expertise adds to the substantial time investment for community biolab project work. 

\begin{quote}
    \Tom: \textit{Expertise, I think is the issue. There are people out there who want to get into a project, and the problem is they don't really know how to work a Pipetman, or don't know how to streak a plate or do a PCR. They've heard about all this stuff, but getting them proficient at just some of the basic things you do. I found out this is a lot harder than you would think to get somebody to actually do an experiment. It doesn't come easy to people, apparently. I mean, I've been doing this my whole life, so it's all second nature. But someone coming into this with an interesting idea who has no experience in a lab, I don't think they really realize that it's very different from writing a computer program or doing an electronic circuit. It's a total different skillset that you've gotta learn.}
\end{quote}

\subsubsection{Lab members are coordinating cooperative work with limited software support}

Interviewees shared that they often collaborate with other lab members. Sharing expertise, dividing labor, and the positive social experience were all raised as motivations for group as opposed to solo work. 
Group work was described as being complicated by limited and variable availability of the time of project members, set against the time demands and sensitivities of biowork. And as explored, expertise differentials created frustrations. 
We asked about the software tools lab members relied on to support cooperative work, particularly in the context of coordinating group tasks and sharing project data.  We found that project members were largely relying on generic workplace communication and file management tools, such as email, Slack, and Google Suite, and were using these to communicate technical project information, share expertise, organize project data, schedule work, manage volunteers and manage inventory. In most cases this wasn't framed as a problem, though two participants did indicate that a lack of appropriate tools to coordinate group work was a limitation.

\begin{quote}
    \Wendy: \textit{Communication across members is very difficult. We have a Slack channel and it's often individuals will slack each other about, `I'm coming into the lab on Thursday and I need help with so-and-so'... They might have a laboratory notebook, although nobody is really using a digital lab notebook. So it's a lot of on paper, in-person communication, with Slack being used to mainly coordinate timing of things... I think it's a struggle. I think that a lot of people in our group are working in either silos or very small groups, because there's no clear cut way that seems to be a good way to communicate both data and all of that other information.}
\end{quote}

\begin{figure*}[htbp]
\centering
\includegraphics[width=1.0\linewidth]{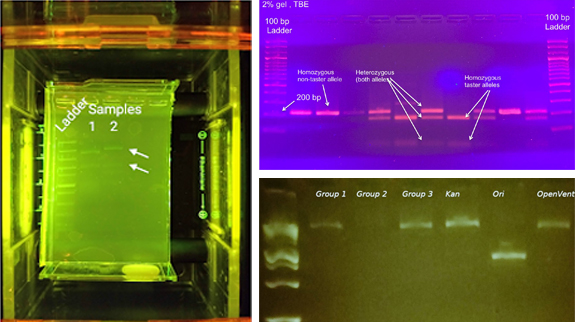}
\caption{DNA fragment analysis through the gel electrophoresis method is an example of the direct visual data analysis frequently performed in community biolabs. This process of separating and sorting stained DNA fragments, alongside a "ladder" of fragments of known length, results in an artifact with lines that glow under UV or blue light (depending on the stain used). Shown here are three such artifacts from labs in our sample, the photos of which were digitally labeled after the fact for illustrative purposes. The practitioner, relying on expertise, infers meaningful information about the DNA signatures of the samples by examining the lines in each column and comparing their positions relative to the ladder and to each other. This image might be stored in a lab notebook, or results might be simply communicated to project members as ‘success’ or ‘failure’ for each of the samples. Analysis of DNA gels was mentioned by multiple interviewees, but is also broadly representative of data analysis methods common in community biolabs that require a careful examination of a physical or digital object to produce a clear, qualitative judgment without the need for computational quantitative analysis, in contrast to `big data' as mentioned in 4.2.4.
}
\Description[Three examples of gel electrophoresis data artifacts with explanatory labels superimposed on the gel images.]{Three photos of gel electrophoresis artifacts, each of which is a rectangle of solid gel with columns of glowing lines. On each gel, a "ladder" line is on one or both sides, with other columns of lines next to it for qualitative comparison. All three images have digital labels superimposed to identify the fragments being analyzed.}
\label{fig:gels}
\end{figure*}

Laboratory biology projects often involve complex, multi-day, timing-sensitive workflows, processing a large number of separate biological samples. Software tools known as Laboratory Information Management Systems (LIMS) are used to track inventory and associated data through workflows in many professional labs, particularly in commercial and biomedical settings. Some interviewees used free software, namely Quartzy and Benchling, to track inventory and associated data. Systems had been set up in two of the labs in our study using QR codes to track inventory for the whole lab or for a specific project, as a means to gain control over the perceived tendency towards disorder inherent to biowork:
\begin{quote}
    \Lisa: \textit{One thing that was really helpful was putting QR codes on everything...We had a little kind of QR code maker. And so every sample got a QR code. Every primer that went into a reaction had a QR code, every PCR product that was produced got a QR code and then an Excel spreadsheet. ... I think we had to kind of develop the systems as we went... because, you know, we lost things and in the typical lab way, we had random tubes labeled 1 through 12 that we were like, What are these?}
\end{quote}

Most participants didn't express a desire for more software tools to support cooperative work. However, considering the current reliance on a limited range of generic tools, there seem to be opportunities for the development of new software options to streamline communication, track inventory and sample data, and otherwise facilitate the interconnected actions of DIYbio practitioners as they design and execute project work.

\subsubsection{Little engagement with `big data'} 
Over the last decade, computationally intensive work requiring the collection and analysis of large datasets (`big data') has become increasingly commonplace within laboratory biology \cite{Fillinger2019}. We found that community biolabs resembled professional biology labs in the equipment, materials, and many of the techniques used, but in stark contrast few projects required any sort of computational data analysis and largely involved work with small numbers of samples. 
Three participants described project work that required computational analysis while eight explained that the data they engaged with during project work could be easily interpreted by eye without computational analysis and in most cases the data were images that could be qualitatively assessed. Data analysis was important to much of the work described to us and significant domain expertise is required to extract meaning from data artifacts in community biolabs. However, data analysis was often restricted to direct visual assessment of the physical or digital products of a machine or lab tool (as in the DNA fragment analysis through the gel electrophoresis workflow shown in Figure \ref{fig:gels}) without the need for data to be downloaded and processed in a digital format. Lab users described collecting data to the extent that this could inform their next physical actions, e.g., visually assessing whether bacteria had grown on an antibiotic containing plate to infer whether a genetic transformation was successful and the new strain could be used in the next phase of the workflow.
A notable exception to this finding was the Barcode the Harbor project at \BUGSS, which relied heavily on computational analysis of large datasets.

Computational data analysis often accompanies laboratory biology work in professional settings, but requires different skill sets as reflected in the emergence of ``bioinformatician'' as a distinct role in some labs \cite{Chang2015}. Three interviewees had professional training in computer programming, another described being \emph{``into bioinformatics''}. Five participants mentioned other lab members with backgrounds in software engineering, computer science, IT, data science, or bioinformatics. 
One lab manager pointed out that many of the people who express interest in their lab are from computer science or bioinformatics backgrounds, and that these people would benefit from more ways to contribute their computational skills to working in the lab. 
Our data suggest that big data work  (`big data') is rare in community biolabs for reasons other than a lack of prerequisite expertise among lab users. Cost could be a reason. Laboratory workflows that generate large volumes of quantitative data are likely to require large amounts of consumable materials such as purified enzymes, which remain prohibitively costly for many practitioners. Coordinating the laboratory work for data generation may be a major barrier for the reasons explored above. However, it could also be that users of community biolabs are not particularly interested or motivated by projects that require extensive time spent processing data away from the lab bench. Further research would be required to further probe the reasons for our finding that big data projects are rare at community biolabs.

\subsubsection{Community biolab members find their work meaningful and rewarding }
We found that motivations for community biolab users were diverse, but that the satisfaction derived from having the freedom to pursue their own curiosity and to learn new skills was mentioned many times. One participant described the ``open learning culture'' that they found in their community biolab, allowing them to develop their skills while engaging in diverse scientific activities, from \emph{``making bioplastics''} to \emph{``teaching 10 year olds what lightning is''}. Beyond the ability to explore curiosity and to learn, many interviewees felt motivated by the idea that their work in the community biolab was contributing to bringing about a broad social change or demonstrating the feasibility of alternative technoscientific futures, with the community biolab acting as stage or microcosm. 
Many of the specific projects described to us were intended to create a resource such as a plasmid (portable form of DNA) or lab protocol (like a recipe) that could be shared with the broader DIYbio community. 
Social change goals were themselves diverse, though themes included creating cooperative communities, broadening \emph{``access to tools of discovery''}, promoting \emph{``curiosity driven research''} and challenging corporate control of the production and distribution of technoscientific products such as pharmaceuticals. 
Participants could not always articulate a specific outcome motivating their work but nevertheless felt that it provided something valuable to themselves and others.

\begin{quote}
    \Tim: \textit{In industry, and I think in academia as well, we tend to think of the merit of a technical project through its... economic output or its scientific output, you know, does it have a novelty to it or something like that? And one thing I really like about this project is it really has, I think, merit almost on a community or an emotional or a spiritual [level]...There's a way in which it can just feel good and fun and it brings people together and people get excited, and that's something that I've really keyed into that I think has made me stick with it this far.}
\end{quote}

\section{Discussion}
Building on our findings, we discuss the factors that set technosicentific work in community biolabs apart from either professional biolabs or makerspaces, and consider some of the pressing challenges and research opportunities for system design to support work in community biolabs. Surprisingly, we find that the study and design of support systems for cooperative work may be a more fruitful area of work for HCI researchers than the design of open hardware and labware, though much could be learned by studying how community biolab users are innovating in that domain. 

\subsection{Community biolab members are creating new approaches to laboratory biology work}

We are interested in community biolabs as spaces for those looking to expropriate the tools of professional laboratory biology.  We found that projects being pursued by lab members were diverse, spanning projects that could be described as `synthetic biology', `environmental monitoring,' through to `bioart'.  
The project work described to us was not primarily focused around producing large, reliable datasets or peer-reviewed research papers---the typical legitimized products of professional research efforts.  
Nevertheless, we found that much work in community biolabs is making use of professional expertise, commercial lab equipment, and contemporary laboratory biology techniques, putting these to work according to members' own personal motivations and interests. This work is providing practitioners with opportunities to find meaning through learning new skills, creative pursuit of their own curiosity about the world, social connection, and a sense of valuable social contribution. 

Project work in community biolabs is driven by the interests of project members. 
While the equipment, materials and techniques being employed in community biolabs in many ways resemble those of a professional biology laboratory, the direction and management of project work perhaps more closely resembles work in makerspaces, with individuals or groups setting their own agendas and timetables to pursue interests that may have little in common with other users in the same space beyond a shared reliance on certain pieces of equipment. This is not to say that community biolab work is not performed within a social context that will direct to some extent the selection of projects and the manner of execution. 
Most notably, every lab implemented some sort of safety screening system that would prohibit work with certain organisms or materials. In addition, even for projects that are not intentionally collaborative, social interaction with other lab members is likely unavoidable. In most cases labs were described as being space-constrained, necessitating a degree of negotiated usage, exacerbated by the long time scales of many life science projects.  Several examples were shared with us of chance interactions and sharing of skills or ideas leading to new directions for project work. However, a clear contrast can be made against the university life science research laboratory, in which the creative freedom of research personnel must operate in accordance with norms, standards and priorities generated within a hierarchical social structure.  Future research would be needed to reveal the likely more subtle and fluid ways that the social structures and dynamics of community biolabs influence the work within them. 

Challenging norms within professional technoscience was specifically called out by some participants as a motivation, and at some level, community biolabs are inherently a critique of the gate-kept silos of professional science systems. We can see community biolabs as  \textit{laboratories} for new approaches to doing  biowork. However, further research should be carried out to examine in more detail how community biolabs accommodate technically demanding work without relying on the more rigid infrastructures established within professional labs to support and direct work. 

Makerspaces and community biolabs resemble one another in many respects as accessible, communal sites for hands-on creative and technical project work. Structured educational activities and events proceed alongside private project work. Such is the synergy that many community biolabs are co-located with or housed within makerspaces. Work in community biolabs seems to resemble work in makerspaces or hackerspaces not just for its sociality but also in motivations and worldview, with many members feeling that they are giving back to some broader community or driven by a sense of challenging dominant technoscientific or economic norms. However, work with organisms and biomaterials requires not only a unique set of domain specific knowledge and skills, but also a different rhythm of work with even a single workflow spanning days or weeks, with few opportunities to assess whether a given step has been successful or whether instead expensive or irreplaceable materials have gone to waste. Work in community biolabs brings the skills and mindset of professional life science work to a social context like that of a makerspace. 
This highlights the need for HCI research into systems that can support cooperative and non-hierarchical work while accounting for the particularities of laboratory science practice.

\subsection{Moving beyond facsimiles of commercial lab equipment}

Building low-cost equivalents of commercially available lab equipment, such as centrifuges and thermocyclers, was viewed as an important enabling activity by the founders of DIYbio.org \cite{Mac_Cowell2009-er,Tocchetti2014-bh}. Our findings suggest that for the set of community biolabs we researched, building low-cost equivalents of commercial lab equipment is a reasonably common activity but is not a prerequisite or dominant concern for work in community biolabs.  
Quality lab equipment is easily available at no- or low-cost via donations or purchase through second hand re-sellers, though this may only apply to the community biolabs located in metropolitan areas with large universities and biotech industries.

Making equipment was not necessarily an important way to save money, but was still being pursued by many users of community biolabs. Motivations for this work included the building process itself offering a chance to learn new skills, and understanding the basic mechanisms underlying equipment of that type. Indirectly, equipment making seemed to be an opportunity for lab members to form new social interactions either by assisting in the process of making or appreciating the spectacle of the making and the artifact. Another possible motivation, only hinted at, was that equipment can be made to desired measurements, particularly relevant for accommodating bulky equipment into non-standard spaces within already overfilled labs. Our findings demonstrate that saving money at point of purchase is only one possible motivation for DIY equipment making and likely the least relevant for community biolabs. 

\citet{Chagas2018} has explored the benefits of open source science hardware for the ``haves'' and the ``have nots''. However, we found it hard to accommodate community biolabs within a dichotomy that separates the resource-plentiful and the very resource-constrained. Community biolabs present a potentially unique setting for research into the design and adoption of open source laboratory hardware, simultaneously resource-rich in many ways (equipment, expertise, dedicated space) and yet also constrained (equipment not well maintained, expertise not uniform, limited space). In the domain of DIY tool making, we found a diversity of experiences and expertise as well. For some lab users, designing a piece of lab equipment may be their primary focus, whereas for others this is a peripheral interest. 
\citet{Fernando2020-bq} described an open science workshop carried out at a community biolab in New York, USA, finding exciting implications for the dissemination of open science hardware internationally. We suggest that such approaches, or longer term participatory design work housed within community biolabs, might also be a fruitful method for HCI researchers to learn about the unique needs, challenges, and affordances for DIY tool making in these hybrid settings of resource abundance and constraint.

\subsection{Identifying research opportunities to support cooperative work in community biolabs }
 
Community biolab members looking to pursue cooperative project work face a number of concurring challenges. The two challenges that emerged most clearly from our findings are time constraints and skill barriers. 
As explored above, much laboratory biology work is both time-consuming (long protocols or long incubation times) and time-sensitive (organisms must be grown for specific lengths of time). 
Users of community biolabs are almost always hobbyists, with community biolab work a minor activity in their lives in terms of time allotment. The travel times to the few lab spaces serving large metropolitan regions further exacerbate the time constraint challenge for cooperative work in community biolabs. 
Under these conditions, it is clear that coordinating teams of lab users to work cooperatively on tasks in person is a significant challenge, as is simply finding sufficient time to communicate as a team about goals, work, and results even if this work is done online and remotely. 

The other major challenge that participants raised was that potential collaborators often lack the skills necessarily to meaningfully contribute without creating substantial additional labor for the more experienced scientist leading the project. Almost all of the participants in our study were experts with substantial formal training or years of first-hand work experience in a relevant STEM field. Nevertheless, due to the extreme specialization within STEM professions, many lab members described trying to accomplish challenging tasks for which they had no prior training. 
Many talked about their struggles to use their very limited time to train potential collaborators who are novices with respect to laboratory biology to the point where they could become productive team members. 
They hinted at a diverse set of technical and soft skills that were found to be lacking in some potential collaborators, ranging from patience and failure-tolerance through correct technique for setting up a PCR reaction. Further data gathered specifically from community biolab users who do not have a prior formal training in biology would be useful to establish whether the skill and knowledge barriers perceived by expert lab members also lead to novice users feeling that their work is unsatisfying or unproductive. We can at least confidently say that expert community biolab members would like to have better tools or approaches for training novice collaborators. 

Using the kernel metaphor of research infrastructure \cite{10.1145/2531602.2531700}, our study begins to identify likely common elements of the community biolab \textit{cache} (resources such as equipment and procurement services) and the \textit{addressing} (work to make cache resources available, in which paid or volunteer lab managers play a leading role). Future work would be needed to elucidate these elements further. 
Digital tools and platforms to support the addressing of research infrastructures are increasingly common for cooperative work in professional life sciences. We found that project work in community biolabs is being supported by generic software tools for communication and file management, such as Google Suite and Slack, as well as platforms for data and inventory management such as Benchling and Quartzy. A digital platform that allowed community biolab users to access and communicate information about lab resources including equipment, materials, researcher schedules, and datasets might go some way to alleviating some of the challenges faced by community biolab users. However, it is unclear whether the value provided by such a resource would be sufficiently compelling to community biolab users to justify the required level of coordinated action to create and maintain it.
Therefore, we believe that HCI researchers could play a key role in establishing systems that support non-traditional and cooperative technoscientific work.

\section{Conclusion}
In this research, we set out to understand the contemporary realities of DIYbio through community biolabs as shared spaces of practice. 
By studying 11 different community biolabs, we found that lab members engaged in a variety of laboratory biology projects using shared infrastructure and community vetting of project safety.
These projects ranged in focus and method, and included bioart and work on local environmental concerns.
We found that community biolabs are well-supplied to take on these projects with donated or cheaply sourced second-hand equipment from professional life science laboratories. 
We encountered a few examples of lab users building DIY equipment for a range of reasons including the educational value of fabrication activities, as well as the novel possibilities afforded by custom equipment.
However, we did not find DIY equipment to be a key enabling activity in the community biolabs.
Despite the differences in project materials, equipment, goals, and motivations,  the community biolab users and managers that we interviewed were nearly unanimous in their insistence that no alternative places existed where they could engage in this work. 
The main constraints we found that lab members faced with respect to project work had more to do with collaboration challenges, time constraints and the realities of working with living organisms, rather than lack of equipment or cost.
Based on our findings, we argue that community biolabs are vital, dynamic,  well-resourced places where DIYbio practices are being developed and deployed every day. 
We believe that these sites offer an alternative to the\textit{ status quo }of life science laboratories. 
In community biolabs, cutting-edge bioscience equipment, materials, and expertise are found free from constraints of the professional laboratory. 
Therefore, we are hopeful that community biolabs are places where we might ``stay with the trouble'' \cite{Haraway2019Tools}. 
Based on our research, we believe that HCI can support these sites of practice by considering the design of systems that can support ad hoc collectives pursuing complex and time-sensitive work for knowledge-making.
https://www.overleaf.com/project/624f35f8898fd7021d7513f0

\begin{acks}
We would like to thank all participants in this study both for generously offering their time to participate and for the work they do to imagine and build new systems of biotechnoscience framed around access and altruism. 
We especially thank community biolabs who participated: Biotech Without Borders (\href{https://biotechwithoutborders.org/}{biotechwithoutborders.org}), 
Boslab (\href{https://www.boslab.org/}{boslab.org}), 
BUGSS (\href{https://bugssonline.org/}{bugssonline.org}), 
Cambridge Biomakespace (\href{https://biomake.space/home}{biomake.space}), 
Counter Culture Labs (\href{https://www.counterculturelabs.org/}{counterculturelabs.org}), 
Hive Biolabs (\href{https://kumasihive.com/hive-biolab/}{kumasihive.com/hive-biolab}), 
London Biohackspace (\href{https://biohackspace.org/}{biohackspace.org}), 
Open Science Network (\href{https://www.opensciencenet.org/}{opensciencenet.org}), 
ReaGent (\href{https://reagentlab.org/en/}{reagentlab.org}), 
SoundBio (\href{https://www.sound.bio/}{sound.bio}), 
Spira, Inc. (\href{https://www.spirainc.com/}{spirainc.com}), 
TriDIYBio (\href{http://www.tridiybio.org/}{tridiybio.org}).
Cartoon graphics used in figures were created using BioRender.com. 
We thank the Gordon and Betty Moore Foundation for supporting this research.
\end{acks}

\bibliographystyle{ACM-Reference-Format}
\bibliography{DIYbioReferences}


\end{document}